\documentclass[aps,prb,twocolumn]{revtex4-2}
\usepackage{epsfig,amsopn,siunitx}
\usepackage{graphicx}
\usepackage{array}
\usepackage{color}
\usepackage{bm}
\usepackage{hyperref}
\hypersetup{
    colorlinks=true,
    linkcolor=magenta,
    citecolor=magenta,
    urlcolor=magenta
}
\usepackage{amsmath,amssymb}
\usepackage{enumerate}
\newcommand\bea{\begin{eqnarray}}
\newcommand\eea{\end{eqnarray}}
\newcommand\beq{\begin{equation}}
\newcommand\eeq{\end{equation}}

\def\nn{\nonumber}

\def\ep{\epsilon}

\def\si{\sigma}

\def\De{\Delta}

\def\ua{\uparrow}
\def\da{\downarrow}

\begin{document}

\title{Topologically protected zero-reflection nodes for perfect crossed Andreev reflection}
\author{Abhiram Soori} 
\email{abhirams@uohyd.ac.in}
\affiliation{School of Physics, University of Hyderabad, Prof. C. R. Rao Road, Gachibowli, Hyderabad 500046, India}

\begin{abstract}
Generating non-locally entangled electron pairs via Cooper-pair splitting is vital for solid-state quantum information processing. However, isolating the underlying crossed Andreev reflection (CAR) is challenging due to competing transport processes like electron tunneling (ET) and local Andreev reflection (AR). Here, we propose a flux-tunable four-terminal normal metal-superconductor junction that achieves deterministic, 100\% efficient CAR. We demonstrate that at exactly half a magnetic flux quantum ($\phi=\pi$), exact destructive Aharonov-Bohm and Peierls interferences structurally forbid ET and AR, respectively. By tuning the central junction hopping, electron reflection is also suppressed to zero. Using the Cauchy argument principle, we prove that this zero-reflection node is topologically protected by a quantized winding number in parameter space. When combined with the precise spatial symmetries of the junction, this exact suppression guarantees a unit CAR probability. We establish that this regime is characterized by a strictly positive cross-correlation shot noise, providing an unambiguous experimental signature of Cooper-pair splitting. Furthermore, this perfect CAR is nearly broadband within the superconducting gap. 
While generic structural disorder breaks the spatial symmetries required for the exact cancellation of ET and AR, the crossed Andreev reflection--though no longer perfectly quantized --remains highly dominant under weak disorder, offering a resilient architecture for deterministic nonlocal entanglement generation.
\end{abstract}

\maketitle
\section{Introduction}
 In normal metal-superconductor-normal metal (NM-SC-NM) heterostructures, non-local phase-coherent transport involves crossed Andreev reflection (CAR) -- a process that non-locally converts an incoming  electron from one terminal into an outgoing hole in another~\cite{recher2001,beckmann2004}. The time-reversed analog of this process is Cooper-pair splitting, wherein a voltage bias is applied between the SC and the two NMs and Cooper pairs split into distinct NM leads, generating non-locally entangled electron pairs. These electron pairs serve as an essential resource for quantum communication~\cite{horodecki2009}. 

However, enhancing CAR is notoriously difficult due to the ubiquitous presence of competing transport mechanisms, most notably electron tunneling (ET), alongside local Andreev reflection (AR) and electron reflection (ER). Experimentally, CAR is identified by a negative nonlocal differential conductance, $G_{RL}$. Because CAR is heavily masked by ET, that contributes positively to $G_{RL}$, the net nonlocal conductance in standard, realistic setups is typically positive.

To overcome this and favor CAR, several strategies have been proposed~\cite{smitha2002,Chtchel2003,yamashita2003,melin2004,hankie2013,soori2017,manisha2018,fuchs2021}, and some of these have been experimentally realized~\cite{russo2005,beckman2004,Feng2025}. Notable approaches include introducing potential barriers at the NM-SC interfaces with a separation comparable to the superconducting coherence length~\cite{Chtchel2003,russo2005}, and employing anti-parallel spin-polarized ferromagnetic leads, which  suppress ET due to  spin-conservation~\cite{yamashita2003,melin2004,beckman2004}. 

Achieving {\it perfect} CAR (with unit probability) is of paramount importance, as it corresponds to deterministic Cooper-pair splitting efficiency. Recent theoretical proposals to reach this ideal limit have predominantly invoked exotic materials, such as topological superconductors coupled to quantum anomalous Hall insulators~\cite{Zhang2017}, spin-orbit coupled graphene junctions~\cite{zhao2023}, and Dirac semimetals~\cite{trauzettel2019}. In four-terminal geometries consisting  topological insulators, perfect CAR has been predicted to arise from constructive interference under specific parameter regimes~\cite{wei2020}. 

\begin{figure}
\includegraphics[width=\columnwidth]{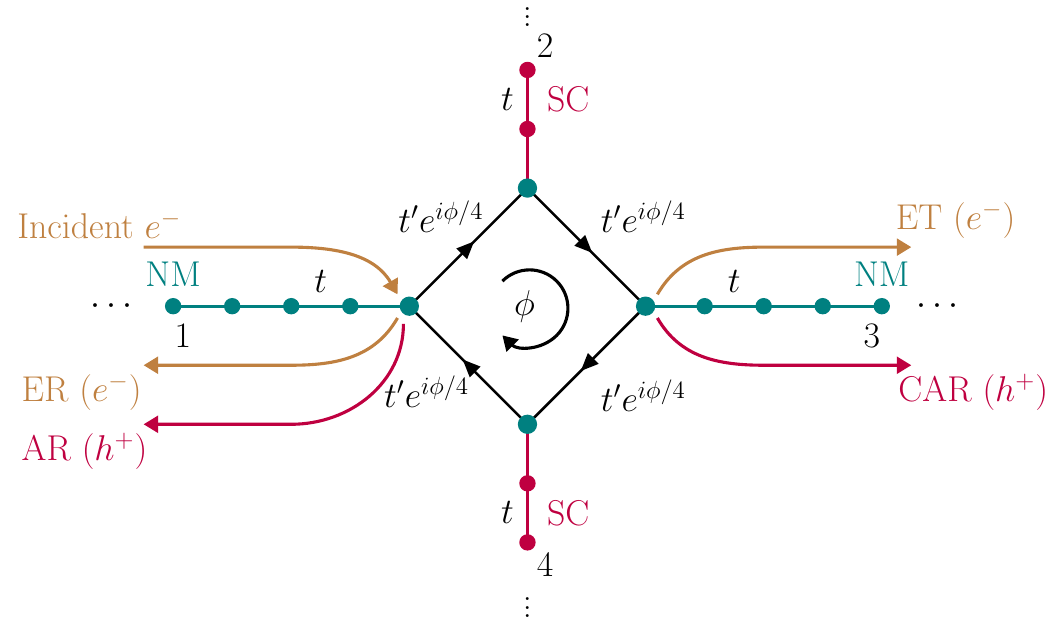}
\caption{Schematic of the flux-tuned junction. Two normal wires (1, 3) and two superconducting wires (2, 4) are coupled via a central square lattice threaded by a $\phi$-flux. For an electron incident from wire 1, destructive interference at $\phi=\pi$ completely suppresses both ET and AR. Tuning the lattice hopping to $t' \simeq t/\sqrt{2}$ additionally nullifies ER, enabling perfect CAR.}\label{fig:schem}
\end{figure}

In this paper, we propose a four-terminal junction comprising two normal metal and two superconducting leads coupled via a central square lattice threaded by a magnetic flux $\phi$ [see Fig.~\ref{fig:schem}]. 
We demonstrate that at half a flux quantum ($\phi=\pi$), the system exhibits perfect CAR, protected by {\it topological winding} of the phase of ER amplitude in parameter space and arising from exact, destructive Aharonov-Bohm and Peierls interferences that structurally forbid ET and AR, respectively.

Specifically, threading the central square lattice with exactly half a magnetic flux quantum ($\phi=\pi$) structurally forbids both ET and local AR through exact Aharonov-Bohm and Peierls destructive interferences. With ET and AR eliminated, the scattering dynamics reduce to a strict competition between CAR and ER. We show that for a specific junction hopping amplitude, the ER amplitude is exactly zero, yielding a deterministic CAR probability of unity. Dictated by the Cauchy argument principle~\cite{churchillnbrown}, this complete suppression of ER is topologically protected, manifesting as a nonzero winding of the reflection phase around a closed loop in a two-dimensional parameter space.

The remainder of this paper is organized as follows. Section~\ref{sec:calc} details the theoretical model and calculational framework. Section~\ref{sec:res} presents the numerical results alongside their physical analysis. In Section~\ref{sec:expt}, we discuss the experimental feasibility of the proposed architecture. Finally, Section~\ref{sec:con} provides our concluding remarks, while Section~\ref{sec:app} outlines the detailed derivation of the cross-correlation shot noise formula.

\section{Details of calculation}~\label{sec:calc} 
The full Hamiltonian of the system reads $H = H_1 + H_2 + H_3 + H_4 + H_c$, where
\begin{align}
H_m &= 
\begin{cases}
  \begin{aligned}
    &\sum_{n=0}^{\infty} ~[-t(\Psi^{\dag}_{m,n+1}\tau_z\Psi_{m,n} + {\rm H.c.}) \\
    &\quad ~~-\mu\Psi^{\dag}_{m,n}\tau_z \Psi_{m,n} ], {\rm ~~for}~~m=1,3,
  \end{aligned} &  \\
  \begin{aligned}
    &\sum_{n=0}^{\infty} ~[-t(\Psi^{\dag}_{m,n+1}\tau_z\Psi_{m,n} + {\rm H.c.}) \\
    &\quad ~~-\mu\Psi^{\dag}_{m,n}\tau_z \Psi_{m,n}] -\ep_0 \Psi^{\dag}_{m,0}\tau_z\Psi_{m,0} \\
    &\quad ~~+\De\sum_{n=1}^{\infty}\Psi^{\dag}_{m,n}\tau_y\si_y\Psi_{m,n}, ~~ {\rm for}~~m=2,4,
  \end{aligned} &
\end{cases} \notag \\[2ex]
H_c &= -t'\sum_{m=1}^{4} ~[\Psi^{\dag}_{m+1,0}M_{\phi}\Psi_{m,0}+{\rm H.c.}], ~~ {\rm with} ~5\equiv 1. \label{eq:H}
\end{align}

Here, $m$ and $n$ denote the wire and site indices, respectively. The state at each site is described by the Bogoliubov-de Gennes spinor $\Psi_{m,n}=[c_{m,n,\ua}, c_{m,n,\da}, c^{\dag}_{m,n,\ua}, c^{\dag}_{m,n,\da} ]^T$, where $c_{m,n,\si}$ annihilates an electron with spin $\si$ at site $(m,n)$. The phase accumulation due to the magnetic flux $\phi$ (in units of the flux quantum) threading the central junction is captured by the diagonal matrix $M_{\phi} = \text{diag}(e^{i\phi/4}, e^{i\phi/4}, -e^{-i\phi/4}, -e^{-i\phi/4})$. The real-valued hopping amplitudes within the leads and across the central junction are denoted by $t$ and $t'$, respectively, while $\mu$ is the uniform chemical potential. Finally, $\epsilon_0$ denotes the local onsite energy at the interface site ($n=0$) and $\De$ represents  the superconducting pairing amplitude, in the wires $2$ and $4$. The Pauli matrices $\tau_{z,y}$ act on the particle-hole space and $\sigma_{y}$ acts on the spin space. The matrices that enter the Hamiltonian are given by 
\bea
\tau_z = \begin{pmatrix}
1 & 0 & 0 & 0 \\
0 & 1 & 0 & 0 \\
0 & 0 & -1 & 0 \\
0 & 0 & 0 & -1
\end{pmatrix}, \quad 
\tau_y \sigma_y = \begin{pmatrix}
0 & 0 & 0 & -1 \\
0 & 0 & 1 & 0 \\
0 & 1 & 0 & 0 \\
-1 & 0 & 0 & 0
\end{pmatrix}.
\eea

 Due to spin-degeneracy, the Hamiltonian decouples into two identical, independent spin sectors: (i) spin-up electron with spin-down hole, and (ii) spin-down electron with spin-up hole. Consequently, we evaluate the nonlocal conductance in a single spin sector and include an overall factor of $2$. The corresponding scattering eigenfunction takes the form $\psi_n = [\psi_n^e, \psi_n^h]^T$, and is given by
\bea
\psi_{1,n}^e &=& e^{-i k_e n} +r_{e}e^{i k_e n},~~
\psi_{1,n}^h ~=~ r_{h}e^{-ik_hn} \nn \\ 
\psi_{2,n} &=& \sum_{j=1,2} s_{2,j} [u_j,~v_j]^Te^{ik_jn}, \nn \\ 
\psi_{3,n}^e &=& t_e e^{ik_en}, ~~ \psi_{3,n}^h ~=~ t_h e^{-ik_hn}, \nn \\
\psi_{4,n} &=&  \sum_{j=1,2} s_{4,j} [u_j,~v_j]^Te^{ik_jn},
\eea
where $k_{e(h)} = \cos^{-1}[-(\mu \pm E)/2t]$ and $k_{1(2)} = \cos^{-1}[-(\mu \pm i\sqrt{\Delta^2 - E^2})/2t]$ denote the normal and superconducting lead wavevectors, respectively, with $u_1 = u_2 = \Delta$ and $v_j = E + \mu + 2t\cos k_j$, for $j=1,2$, and $u_j, v_j$ are the electron and hole components of the wavefunction in the superconductor. Matching the scattering wavefunctions via the Schr\"odinger wave equation at the junction uniquely determines the scattering amplitudes $r_e$, $r_h$, $s_{m,j}$, $t_e$, and $t_h$.

 The nonlocal differential conductance between the normal leads is given by  $G_{RL} = \frac{2e^2}{h}(P_{\text{ET}} - P_{\text{CAR}})$, where $P_{\text{ET}}=|t_e|^2$ and $P_{\text{CAR}}=|t_h|^2\sin{k_h}/\sin{k_e}$ denote the probabilities of ET and CAR, respectively, and the factor of $2$ accounts for spin degeneracy. A negative nonlocal conductance ($G_{RL} < 0$) serves as a unambiguous signature of CAR dominance over ET, culminating in the quantized limit $G_{RL} = -2e^2/h$ for ideal, unit-probability CAR ($P_{\text{CAR}}=1, P_{\text{ET}}=0$).

 \begin{figure}[htb]
 \includegraphics[width=4.01cm]{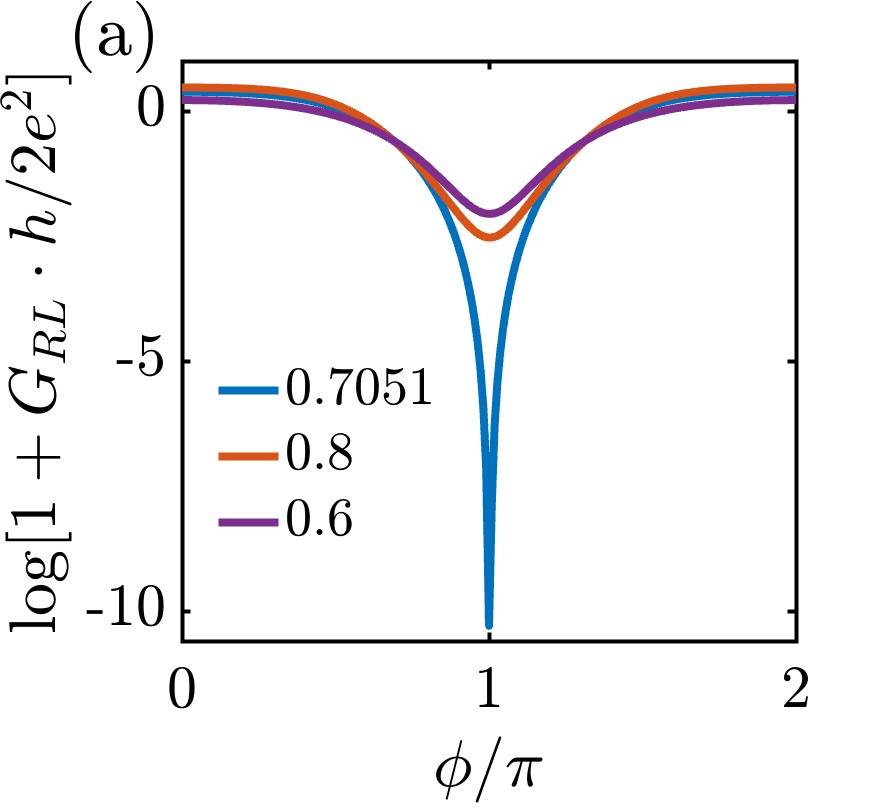}
 \includegraphics[width=4.01cm]{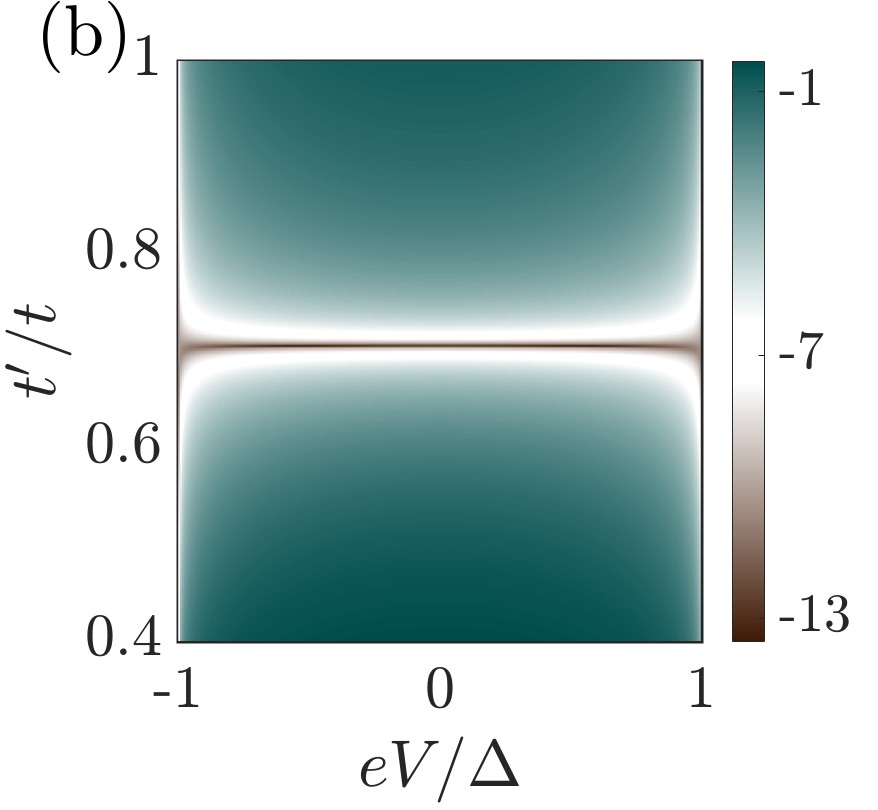} 
\caption{Giant enhancement of  crossed Andreev reflection. (a) The logarithmic nonlocal conductance, $\eta=\log[1+G_{RL} h/(2e^2)]$, versus magnetic flux $\phi$ at zero bias for various junction hoppings $t'/t$ (shown in the legend). (b) Dependence of $\eta$ on bias voltage $V$ and hopping $t'$ at $\phi=\pi$.  Parameters are $\mu=-t$, $\De=0.01t$, and $\ep_0=0.6\De$.}\label{fig:res1}
 \end{figure}
 
\section{Results}~\label{sec:res}
In this section, we present and analyze the main transport results. We begin in Sec.~\ref{sec:aretsupp} by detailing the interference mechanisms that suppress ET and AR to zero. Section~\ref{sec:car} then examines the emergence of perfect CAR, followed by a topological characterization of the zero ER amplitude in Sec.~\ref{sec:topo}. Next, we analyze the cross-correlation shot noise in Sec.~\ref{sec:noise}. Finally, Sec.~\ref{sec:disorder} addresses the impact of spatial disorder on the robustness of CAR.

\subsection{Suppression of AR and ET}~\label{sec:aretsupp}
The physical mechanism that simultaneously suppresses both ET and AR is  rooted in the exact destructive interference induced by the $\pi$ flux. Specifically, ET vanishes because the electron transmission trajectories spanning opposite halves of the square (via terminals 2 and 4) accumulate a relative Aharonov-Bohm phase of exactly $\pi$. Similarly, the AR amplitude is identically zero: an electron incident from terminal 1 may be Andreev-reflected at either of the superconducting terminals (2 or 4). Because the Andreev-reflected hole acquires a conjugate Peierls phase, the total phase accumulated along the $1 \to 2 \to 1$ excursion is $\pi/2$, whereas the $1 \to 4 \to 1$ excursion yields $-\pi/2$. This  $\pi$ phase difference exactly cancels the net AR amplitude. 

With ET and AR structurally forbidden ($P_{\text{ET}} = P_{\text{AR}} = 0$), the conservation of probability reduces to $P_{\text{ER}} + P_{\text{CAR}} = 1$. Consequently, achieving unit CAR probability ($P_{\text{CAR}}=1$) is equivalent to finding a node where the electron reflection amplitude vanishes ($r_e=0$).

\subsection{Perfect CAR}~\label{sec:car}
To highlight the approach toward a unit CAR efficiency, Fig.~\ref{fig:res1}(a) displays  $\eta = \log[1 + G_{RL}h/(2e^2)]$, derived from the zero-bias nonlocal conductance, as a function of the central magnetic flux $\phi$ for various junction couplings $t'$. At $\phi=\pi$ and $t' \approx 0.7051t$, $\eta$ exhibits a pronounced negative minimum, marking the onset of near-perfect CAR. Figure~\ref{fig:res1}(b) maps $\eta$ in the parameter space of incident bias $eV$ and hopping strength $t'$. Crucially, this near-perfect CAR regime is maintained at $t' \simeq 0.7t$ across all bias energies within the superconducting gap ($|eV| < \Delta$), demonstrating its robust broadband nature. When $t'$ deviates from this optimal value, the CAR efficiency remains substantially enhanced as the incident energy approaches the gap boundaries ($E \to \pm \Delta$). This boundary enhancement is driven by a resonant increase in the local Andreev reflection  probability at the superconducting terminals (terminals 2 and 4). As the incident electron energy approaches $\Delta$, the probability of an incident electron being Andreev reflected as a hole at the superconducting interfaces nears unity. This is rooted in diverging local density of states in the superconductor. However, exact Peierls destructive interference structurally forbids this Andreev-reflected hole from backscattering into the incident normal lead (terminal 1). Because the probability of normal electron reflection is tuned to zero at terminals 2 and 4, the incident electron from terminal 1 is perfectly transmitted into terminal 3 as a hole, guaranteeing a highly dominant  CAR.

The physical mechanism underlying the complete suppression of normal electron reflection ($r_e = 0$) at $t' \approx 0.7051t$ can be understood as a wave-mechanical multipath destructive interference phenomenon. When an incident electron reaches the central junction, the total reflection amplitude is a coherent superposition of a directly reflected wave at the interface barrier and a secondary wave that tunnels into the central $4$-site square cavity, circulates through the junction, and is re-emitted back into the incident lead. Tuning the interface hopping parameter $t'$ acts as a barrier transparency knob that precisely adjusts the relative amplitude and phase of this re-emitted wave to achieve total destructive cancellation with the directly reflected wave. 
Notably, this optimal ratio $t' \approx 0.7051t$ is fundamentally a geometric scattering property of the central $4$-site plaquette, closely matching the zero-reflection condition $t' = t/\sqrt{2} \approx 0.7071t$ of a purely normal-metal four-wire junction~\cite{soori2026nwire}. The slight deviation from $t' = t/\sqrt{2}$ is induced by the finite superconducting pairing strength $\Delta = 0.01t$.
  At a threaded magnetic flux of $\phi=\pi$, exact Aharonov-Bohm and Peierls interferences structurally forbid both electron tunneling ($P_{\text{ET}}=0$) and local Andreev reflection ($P_{\text{AR}}=0$). Consequently, closing the normal reflection channel ($P_{\text{ER}} = |r_e|^2 = 0$) via this tuned multipath interference leaves Cooper-pair splitting as the sole remaining scattering channel, strictly forcing $100\%$ of the transport into perfect crossed Andreev reflection ($P_{\text{CAR}}=1$).

\begin{figure}[htb]
\includegraphics[width=4.01cm]{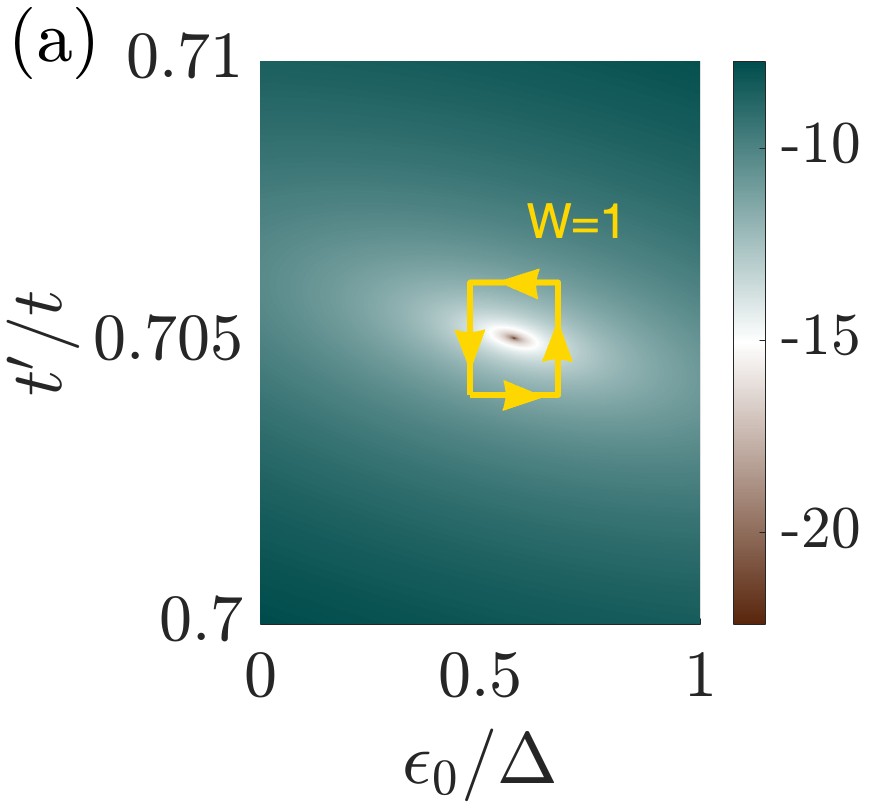}
\includegraphics[width=4.01cm]{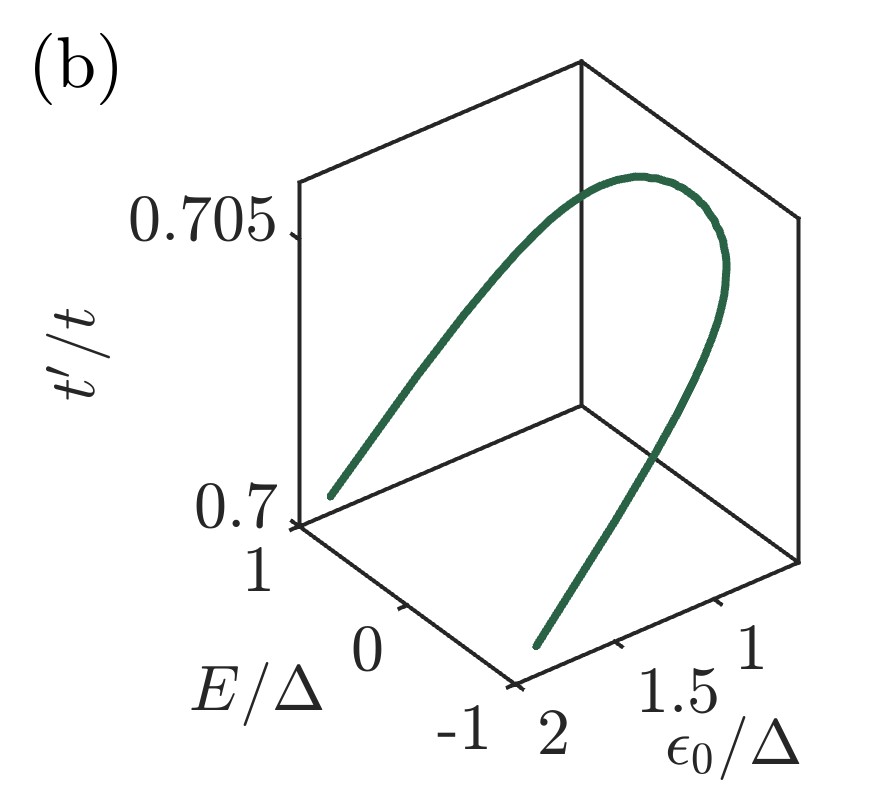}
\caption{Topological protection of perfect CAR. (a) Map of $\eta$ versus $\ep_0$ and $t'$ at zero bias. The phase winding of the electron reflection (ER) amplitude along the indicated closed loop yields $W=1$, topologically guaranteeing a zero of ER (perfect CAR) within the enclosed region. (b) Trajectory of the perfect CAR condition in the $(E,\ep_0,t')$ parameter space. Other parameters are the same as in Fig.~\ref{fig:res1}.}\label{fig:res2}
\end{figure}

\subsection{Topological characterization}~\label{sec:topo}
At $\phi=\pi$, exact destructive interference strictly forbids both AR and ET ($P_{\text{AR}} = P_{\text{ET}} = 0$), reducing the conservation of probability to $P_{\text{ER}} + P_{\text{CAR}} = 1$, where $P_{\text{ER}}=|r_e|^2$ and $P_{\text{AR}}=|r_h|^2\sin{k_h}/\sin{k_e}$. Consequently, achieving unit CAR probability ($P_{\text{CAR}}=1$) is equivalent to finding a node where the electron reflection amplitude vanishes ($r_{{e}}=0$). Because the physical reflection amplitude is a complex quantity, $r_e = \text{Re}(r_e) + i\text{Im}(r_e)$, achieving an exact zero requires both its real and imaginary parts to vanish simultaneously. In the two-dimensional parameter space of onsite energy $\ep_0$ and hopping strength $t'$, the contours satisfying $\text{Re}(r_e) = 0$ and $\text{Im}(r_e) = 0$ intersect at a singular point. Analogous phase singularities have been explored in the context of Brewster reflection in optics~\cite{nirmal2021}.

To demonstrate the topological origin and existence of this zero, in Fig.~\ref{fig:res2}(a) we map $\eta$ in the two-dimensional parameter space of $\ep_0$ and $t'$ at fixed $\phi=\pi$. We compute the phase accumulation of the reflection amplitude $r_{{e}}$ along a closed contour encircling the pronounced minimum of $\eta$ [the yellow loop in Fig.~\ref{fig:res2}(a)]. Methodologically, this topological winding number $W$ is evaluated numerically by discretizing the closed contour into a dense set of $N$ ordered points $\bm{\lambda}_j = (\ep_{0,j}, t'_j)$ and summing the phase differences between adjacent points:
\begin{equation}
W = \frac{1}{2\pi} \sum_{j=1}^{N} \text{Arg}\left[ \frac{r_e(\bm{\lambda}_{j+1})}{r_e(\bm{\lambda}_j)} \right],
\end{equation}
where $\bm{\lambda}_{N+1} = \bm{\lambda}_1$ ensures the loop is closed, and the principal argument function $\text{Arg}(z) \in (-\pi, \pi]$ restricts the phase difference to correctly track the continuous phase evolution by handling branch-cut crossings. Following this integration procedure, we find a quantized topological winding number $W = 1$.

By Cauchy's argument principle~\cite{churchillnbrown}, $W=1$ indicates that the intersection of the zero-contours acts as a topological defect---a phase vortex. This mathematically guarantees the existence of an isolated zero of $r_{{e}}$ within the enclosed region. Crucially, this topological character ensures that the emergence of perfect CAR ($P_{\text{CAR}}=1$) is not a fragile mathematical coincidence but a robust physical inevitability; small generic perturbations to the system may smoothly shift the coordinates of this zero-reflection node in parameter space, but they cannot abruptly destroy it.

In Fig.~\ref{fig:res2}(b), we plot the locus of points in the $(E, \ep_0, t')$-parameter space corresponding to perfect CAR ($P_{\text{CAR}} = 1$). The results show that the values of $t'$ and $\ep_0$ required for unit CAR exhibit only a weak dependence on the incident energy $E$. Furthermore, the parameter window for such $t'$ broadens with an increasing superconducting gap, controlled by the ratio $\De/t$. Interestingly, the optimal value of $t'$ required for perfect CAR in this superconducting architecture is remarkably close to the hopping value that yields a broadband zero-reflection amplitude in a purely NM four-wire junction~\cite{soori2026nwire}, highlighting a fundamental property of the central lattice.

\begin{figure}
\centering
\includegraphics[width=7.5cm]{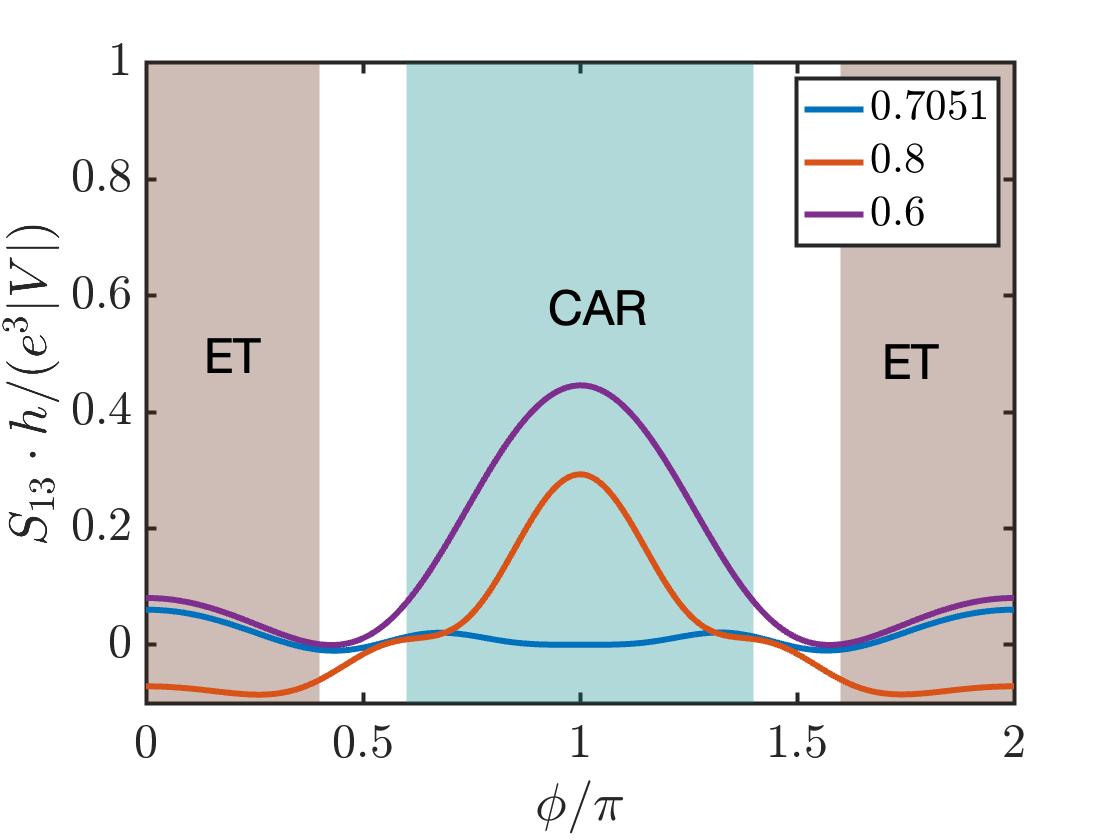}
\caption{\label{fig:noise} 
Cross-correlation shot noise $S_{13}$ between the normal terminals as a function of the applied flux $\phi$ at a small bias voltage $V$ near zero energy for various hopping ratios $t'/t$. The strictly positive noise near $\phi=\pi$ (green shaded region) is a feature of CAR-dominated Cooper-pair splitting. Notably, for the optimal hopping $t'=0.7051t$, the noise vanishes completely at $\phi=\pi$, reflecting the deterministic, fluctuation-free nature of perfect CAR ($P_{\text{CAR}}=1$). Conversely, the brown shaded regions denote the ET-dominated regime. Parameters are the same as in Fig.~\ref{fig:res1}.}
\end{figure}

\subsection{Cross-correlation shot noise }~\label{sec:noise}
To evaluate the nature of non-local transport, we analyze the zero-temperature cross-correlation shot noise $S_{13}$ between normal terminals 1 and 3~\cite{anantram1996,Martin2005}. When a voltage bias $V$ is applied to terminal 1 with all other terminals grounded, $S_{13}$ takes the form [see section~\ref{sec:app} for derivation]
\begin{equation}
S_{13} = -\frac{4e^3|V|}{h} (P_{\text{ET}} - P_{\text{CAR}})(P_{\text{ER}} - P_{\text{AR}}). \label{eq:noise}
\end{equation}

In conventional all-normal-metal mesoscopic junctions, the cross-correlation shot noise $S_{13}$ is negative owing to fermionic antibunching. However, a positive cross-correlation can arise from the competition among the four scattering channels, a feature widely utilized as a hallmark of CAR-mediated Cooper-pair splitting. Previous studies~\cite{melin2010,melin2013,tikhanov2024} have cautioned that a positive $S_{13}$ does not constitute unambiguous evidence for CAR in generic multiterminal geometries, as a combination of local AR and ET can also yield positive cross-correlations.

Indeed, the mathematical structure of Eq.~\eqref{eq:noise} dictates that a positive $S_{13}$ emerges whenever the terms $(P_{\text{ET}} - P_{\text{CAR}})$ and $(P_{\text{ER}} - P_{\text{AR}})$ possess opposite signs. Consequently, positive shot noise occurs not only in the intended CAR-dominated regime ($P_{\text{CAR}} > P_{\text{ET}}$ and $P_{\text{ER}} > P_{\text{AR}}$), but also through a second mechanism: ET-dominated regimes where local AR simultaneously surpasses normal electron reflection ($P_{\text{ET}} > P_{\text{CAR}}$ and $P_{\text{AR}} > P_{\text{ER}}$). As illustrated in Fig.~\ref{fig:noise}, the system enters this latter regime near $\phi=0$ and $2\pi$ for hopping ratios $t'/t=0.7051$ and $0.6$, explaining why $S_{13}$ takes a positive value despite ET dominating over CAR in those regions. Conversely, for $t'/t=0.8$, although ET continues to dominate over CAR near $\phi=0$ and $2\pi$, ER overtakes AR, restoring the conventional negative sign of $S_{13}$.

This ambiguity, however, is completely absent in our proposed ideal operating regime. When the central square lattice is threaded with exactly half a magnetic flux quantum ($\phi=\pi$), exact Aharonov-Bohm and Peierls destructive interferences structurally enforce $P_{\text{ET}} = 0$ and $P_{\text{AR}} = 0$. By structurally eliminating both ET and local AR, non-local current fluctuations arise exclusively from the coherent splitting of Cooper pairs into spatially separated terminals via CAR. Under these specific geometric and magnetic interference conditions, the cross-correlation reduces strictly to a positive quantity proportional to $P_{\text{CAR}}P_{\text{ER}}$. Substituting the probability conservation constraint $P_{\text{ER}} = 1 - P_{\text{CAR}}$ into Eq.~\eqref{eq:noise}, the noise simplifies to a strictly positive value:
\begin{equation}
S_{13}=\frac{4e^3|V|}{h}P_{\mathrm{CAR}}\left(1-P_{\mathrm{CAR}}\right).
\end{equation}
Thus, the strictly positive $S_{13}$ observed at $\phi=\pi$ in Fig.~\ref{fig:noise} provides a direct, unambiguous, and robust signature of dominant crossed Andreev reflection~\cite{Wei2010,Das2012}. Finally, when a symmetric bias is applied to both normal-metal leads relative to the grounded superconducting leads, qualitatively similar features emerge in $S_{13}$, preserving the strictly positive noise signature near $\phi=\pi$.

 \begin{figure}
 \includegraphics[width=4.01cm]{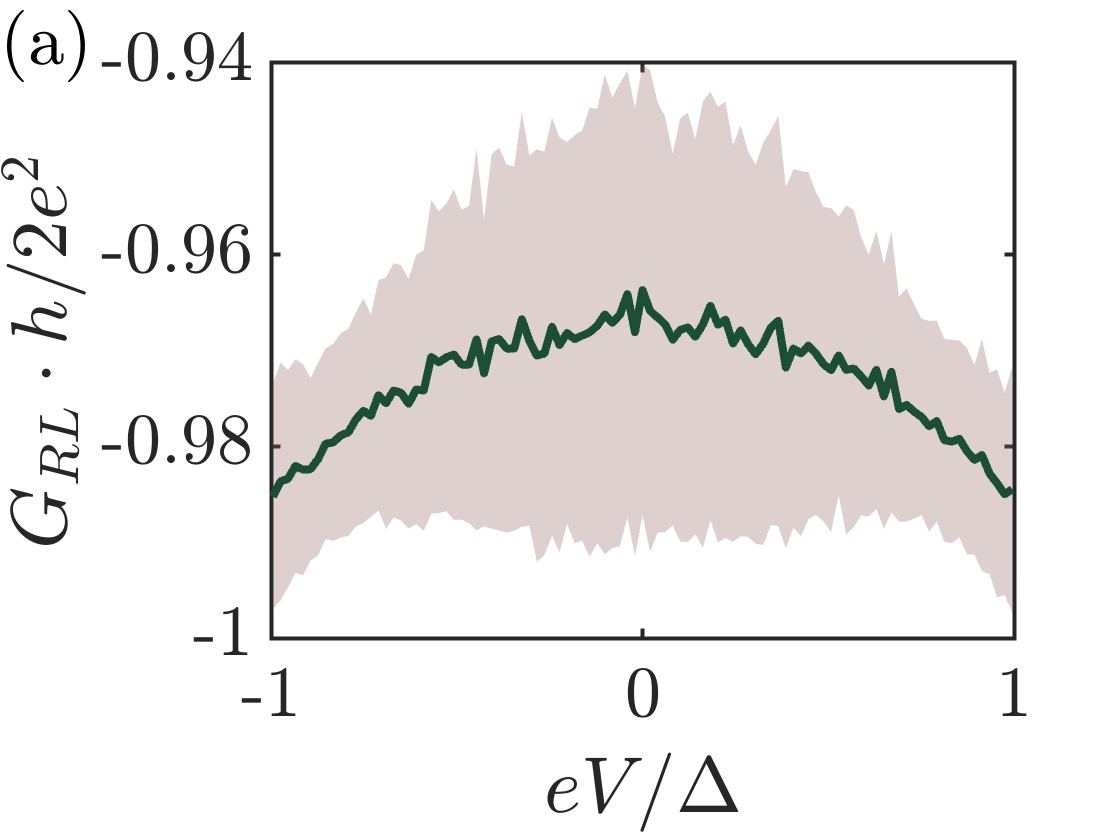}
 \includegraphics[width=4.01cm]{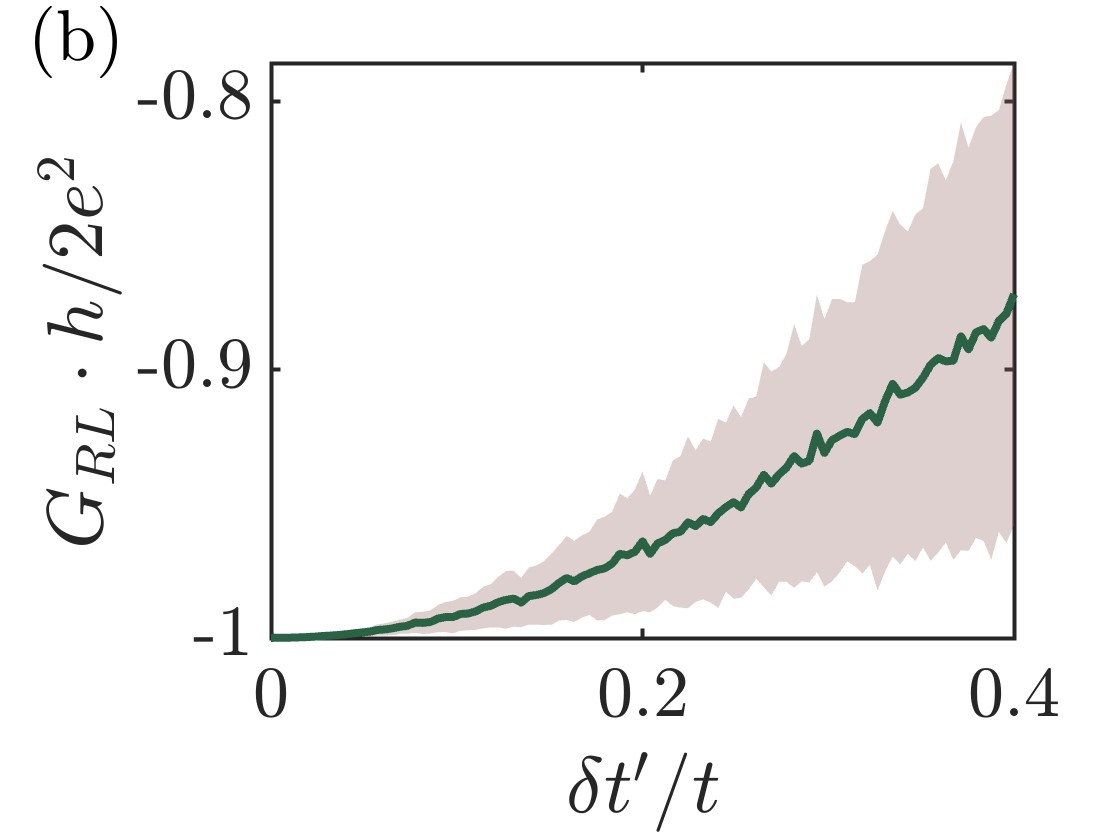} 
 \includegraphics[width=4.01cm]{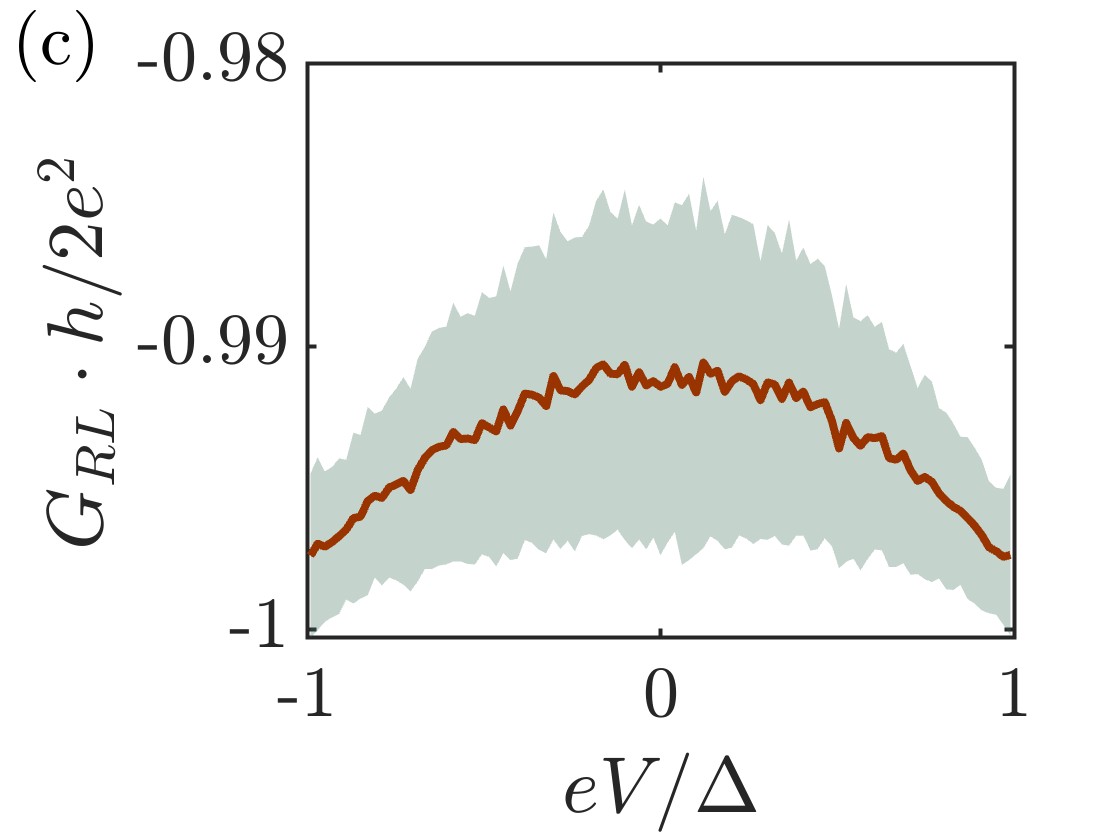}
 \includegraphics[width=4.01cm]{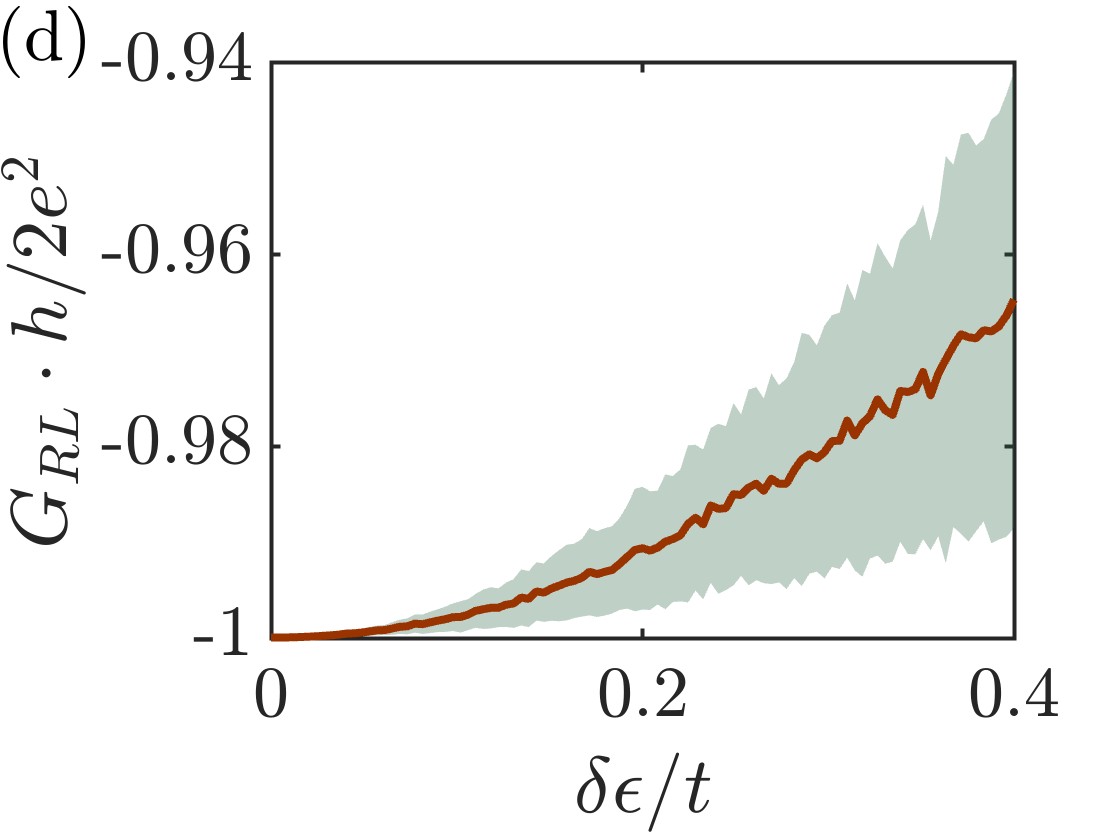} 
 \caption{Robustness of CAR-dominated non-local conductance under structural and chemical potential fluctuations. (a,b) Bond disorder with hoppings in the range $[t' - \delta t'/2, t' + \delta t'/2]$. (a) Bias dependence of $G_{RL}$ at fixed bond disorder amplitude $\delta t'=0.2t$ near $t'=0.706t$. (b) $G_{RL}$ versus bond disorder strength $\delta t'$ at zero bias. (c,d) Onsite disorder with site energies in $[-\delta\ep/2, \delta\ep/2]$. (c) Bias dependence of $G_{RL}$ at fixed onsite disorder amplitude $\delta\ep=0.2t$. (d) $G_{RL}$ versus onsite disorder strength $\delta\ep$ at zero bias. Solid lines denote configuration averages (over 250 realizations) and shaded regions represent standard deviation. The parameters other than $\ep_0$ are identical to those in Fig.~\ref{fig:res1}.}\label{fig:res3}
 \end{figure}
 
 \subsection{Disordered junction}~\label{sec:disorder} 
 In realistic setups, spatial inhomogeneities and fabrication imperfections can break the symmetry of the hopping strengths in the central square. To model this, we incorporate uniform random bond disorder by drawing each central hopping amplitude from $[t' - \delta t'/2, t' + \delta t'/2]$. Figures~\ref{fig:res3}(a) and \ref{fig:res3}(b) display the nonlocal conductance $G_{RL}$ (averaged over $250$ disorder realizations) alongside its standard deviation as functions of bias energy and disorder strength $\delta t'$, respectively. Even for disorder levels as large as $\delta t' = 0.2t$, $G_{RL}$ remains remarkably close to the ideal quantized value of $-2e^2/h$. 
 We subsequently analyze the effect of onsite disorder by choosing the onsite energy at each central junction site (the $n=0$ interface sites connecting the four leads) randomly from $[-\delta\ep/2, \delta\ep/2]$, while keeping the bulk normal-metal and superconducting regions completely pristine.
  Figures~\ref{fig:res3}(c) and \ref{fig:res3}(d) plot the averaged $G_{RL}$ and its standard deviation against bias energy and onsite disorder strength $\delta \ep$ respectively. In both disorder scenarios, while the deviation from ideal transport peaks at zero energy ($eV=0$), it is strongly suppressed near the gap boundaries ($\pm\Delta$) due to the resonant enhancement of local AR at the superconducting terminals. This confirms that the protected interference condition preserves robust non-local entanglement even against moderate structural disorder. 
 
 Notably, generic random disorder destroys the perfect CAR point. The underlying topological protection strictly guarantees only a vanishing ER. Consequently, even at the point in parameter space where the ER amplitude is  exactly zero, the AR and ET amplitudes become finite, rendering the CAR imperfect. Exact cancellation of these ET and AR amplitudes at $\phi=\pi$ requires specific spatial symmetries: (i) identical onsite energies at the SC junction sites (wires 2 and 4), and (ii) symmetric hopping amplitudes across the junction, specifically for bonds $1 \to 2$ ($2 \to 3$) matching those on $4 \to 1$ ($3 \to 4$).
 
 \section{Experimental realization}~\label{sec:expt} 
While our theoretical framework employs a tight-binding lattice model, the proposed architecture maps directly onto state-of-the-art mesoscopic platforms. Highly tunable four-terminal Josephson junctions have recently been realized experimentally~\cite{prosko2024}. To implement the proposed hybrid setup featuring two NM and two SC leads, one could utilize proximitized two-dimensional electron gases, such as semiconductor heterostructures coupled to superconductors~\cite{Wan2015}. In these platforms, the distinct normal and superconducting regions, as well as the central scattering junction, can be precisely defined via selective proximity coupling. Furthermore, threading the required magnetic flux through the central microscopic cavity can be achieved using localized magnetic fields. The practical viability of leveraging such half-quantum magnetic flux ($\phi=\pi$) for topologically protected transport has already been experimentally demonstrated in frustrated Josephson junction nanocircuits~\cite{doucot2002, gladchenko2009}.

In experimental realizations utilizing these proximitized platforms, such as InAs/Al or InSb/Al heterostructures~\cite{bargerbos2020}, the hopping strength $t'$ essentially parameterizes the tunnel barrier between the normal-metal and superconducting regions. This interface coupling can be actively tuned \textit{in situ} through the application of local electrostatic top gates deposited over the junction barriers. By adjusting the applied gate voltages, experimentalists can continuously modulate the height and width of the potential barriers at the NM-SC interfaces, directly changing the effective hopping parameter $t'$. This standard technique provides a highly feasible method for fine-tuning the interface coupling to the specific wave-matching condition ($t' \approx 0.7051t$) necessary for realizing perfect crossed Andreev reflection. 

Together, these advanced fabrication and gating techniques place the realization of our proposed $\pi$-flux architecture, and the subsequent observation of perfect CAR,  within the reach of current experimental capabilities.

  \section{Conclusion}~\label{sec:con} 
We have proposed a conventional mesoscopic architecture that achieves deterministic, perfect CAR without relying on exotic materials. By threading a central four-terminal junction with a half-quantum magnetic flux ($\phi=\pi$), exact destructive Aharonov-Bohm and Peierls interferences structurally forbid the competing transport channels of ET and local AR. We have proven that the complete suppression of the remaining electron reflection is topologically protected by a non-zero winding number in parameter space. Crucially, combining this topologically robust zero-reflection node with the spatial symmetries of the junction yields a unit CAR probability. This perfect non-local entanglement generation is nearly broadband within the superconducting gap. While structural bond disorder at the junction breaks the exact spatial symmetries and introduces slight deviations from the ideal quantized conductance, primarily near mid-gap, the non-local transport signature remains remarkably robust. Finally, utilizing rigorous noise derivations established in multi-terminal junctions involving superconducting leads~\cite{anantram1996}, we demonstrated that the emergence of a strictly positive cross-correlation shot noise provides an unambiguous, fluctuation-free experimental signature of deterministic Cooper-pair splitting. Our findings deliver a highly resilient, experimentally viable blueprint for deterministic entanglement generation, offering a reliable resource for solid-state quantum communication networks.

\begin{acknowledgements}
The author thanks Diptiman Sen and Udit Khanna for valuable and insightful discussions that helped shape and improve this work.
The author acknowledges the use of Google’s Gemini 3.1 Pro AI assistant for assistance in refining the mathematical text, and helpful discussions during the preparation of section~\ref{sec:noise} and Appendix~\ref{sec:app}.  All content was reviewed and verified by the author.
AS acknowledges the financial support received from the Anusandhan National Research Foundation (erstwhile Science and Engineering Research Board) under the Core Research Grant (No. CRG/2022/004311), and the University of Hyderabad. 
\end{acknowledgements}
\appendix
\section{Cross-correlation shot noise}~\label{sec:app}
To derive the shot noise formula for our flux-engineered normal metal-superconductor (NM-SC) junction, we employ the multinomial partition noise approach. This method relies on classical probability laws applied to the mutually exclusive quantum scattering outcomes of independent incident electrons at zero temperature.

\subsection{The Partitioning Concept}
At zero temperature, thermal noise vanishes. Applying a bias voltage $V$ to terminal 1 injects a steady stream of electrons into the junction. Accounting for the 2-fold spin degeneracy, the rate of incident electrons hitting the central junction is $N = 2|eV|/h$. 

Every time a single incident electron hits the central junction, it faces four mutually exclusive scattering outcomes. Let us define a binary occupancy variable $n_i \in \{0,1\}$ for each path:
\begin{itemize}
    \item \textbf{ER ($n_{\text{ER}}$):} Reflects back to terminal 1 as an electron with probability $P_{\text{ER}}$.
    \item \textbf{AR ($n_{\text{AR}}$):} Reflects back to terminal 1 as a hole (Andreev reflection) with probability $P_{\text{AR}}$.
    \item \textbf{ET ($n_{\text{ET}}$):} Transmits to terminal 3 as an electron (electron tunneling) with probability $P_{\text{ET}}$.
    \item \textbf{CAR ($n_{\text{CAR}}$):} Transmits to terminal 3 as a hole (crossed Andreev reflection) with probability $P_{\text{CAR}}$.
\end{itemize}
Probability conservation dictates: $P_{\text{ER}} + P_{\text{AR}} + P_{\text{ET}} + P_{\text{CAR}} = 1$.

Since the outcomes are mutually exclusive, the statistical fluctuations (variances and covariances) of these path occupancies follow a multinomial distribution:
\begin{align}
    \text{var}(n_i) &= P_i(1 - P_i) \\
    \text{cov}(n_i, n_j) &= -P_i P_j \quad \text{(for } i \neq j \text{)}
\end{align}
The negative covariance intrinsically arises because choosing one path explicitly means the particle did \textit{not} choose another (anti-correlation).

\subsection{Cross-Correlation Noise ($S_{13}$)}

We track the net charge $Q_1$ deposited in terminal 1 per incident electron. Defining the elementary charge as $e$, we follow the convention that an electron carries $e$ and a hole carries $-e$.
\begin{itemize}
    \item The incident electron leaves terminal 1: $\Delta Q = -e$.
    \item If ER occurs ($n_{\text{ER}}=1$), an electron returns: $\Delta Q = +e$. (Net $Q_1 = 0$)
    \item If AR occurs ($n_{\text{AR}}=1$), a hole returns: $\Delta Q = -e$. (Net $Q_1 = +2e$)
    \item If ET or CAR occurs, nothing returns to terminal 1: (Net $Q_1 = -e$)
\end{itemize}
The total charge accumulated in terminal 1 is:
\begin{equation}
    Q_1 = -e - e \cdot n_{\text{AR}} + e \cdot n_{\text{ER}}
\end{equation}

We now track the net charge $Q_3$ deposited in terminal 3.
\begin{itemize}
    \item If ET occurs ($n_{\text{ET}}=1$), an electron arrives: $Q_3 = +e$.
    \item If CAR occurs ($n_{\text{CAR}}=1$), a hole arrives: $Q_3 = -e$.
\end{itemize}
The charge equation for terminal 3 is:
\begin{equation}
    Q_3 = e \cdot n_{\text{ET}} - e \cdot n_{\text{CAR}}
\end{equation}
Using the Schottky formula $S = 2 \times \text{rate} \times \text{covariance}$,
the cross-correlation noise $S_{13}$ is proportional to the covariance between the charge fluctuations in terminal 1 and terminal 3:
\bea    S_{13} &=& 2 \left( \frac{2|eV|}{h} \right) \text{cov}(Q_1, Q_3) \nn \\ &=& \frac{4|eV|}{h} \text{cov}(-e \cdot n_{\text{AR}} + e \cdot n_{\text{ER}}, \; e \cdot n_{\text{ET}} - e \cdot n_{\text{CAR}}) \nn \\ &&
\eea
Factoring out $e^2$ and expanding the covariance linearly gives:
\bea   
 S_{13} &=& e^2 \frac{4|eV|}{h} \Big[ -\text{cov}(n_{\text{AR}}, n_{\text{ET}}) + \text{cov}(n_{\text{AR}}, n_{\text{CAR}}) \nn \\ && + \text{cov}(n_{\text{ER}}, n_{\text{ET}}) - \text{cov}(n_{\text{ER}}, n_{\text{CAR}}) \Big] \nn \\ &&
\eea
Applying the multinomial covariance rule $\text{cov}(n_i, n_j) = -P_i P_j$:
\begin{align}
    S_{13}     &= \frac{4e^3|V|}{h} \Big[ P_{\text{AR}}P_{\text{ET}} - P_{\text{AR}}P_{\text{CAR}} \nn \\ & - P_{\text{ER}}P_{\text{ET}} + P_{\text{ER}}P_{\text{CAR}} \Big]
\end{align}
This can also be expressed in the fully factored form:
\begin{equation}
    S_{13} = -\frac{4e^3|V|}{h} (P_{\text{ET}} - P_{\text{CAR}})(P_{\text{ER}} - P_{\text{AR}})
\end{equation}
This formula agrees with Anantram-Datta formula~\cite{anantram1996} for zero temperature cross-correlation shot noise for the case when a small voltage bias $V$ is applied from one NM keeping the SCs and the other NM grounded. 

\bibliography{ref_4wire}
\end{document}